\documentclass[twocolumn,showpacs,preprintnumbers,amsmath,amssymb]{revtex4}
\usepackage{epsfig}

\usepackage{graphicx}
\usepackage{dcolumn}
\usepackage{bm}

\begin{document}

\title{New Constraints on Primordial Minihalo Abundance Using Cosmic Microwave Background Observations}

\author{Yupeng Yang}
\email{yyp@chenwang.nju.edu.cn}
\affiliation{Department of Physics, Nanjing University, Nanjing, 210093, China}
\affiliation{Joint Center for Particle, Nuclear Physics and Cosmology, Nanjing, 210093, China}
\affiliation{National Astronomical Observatories, Chinese Academy of Sciences, Beijing, 100012, China}

\author{Xiaoyuan Huang}
\affiliation{National Astronomical Observatories, Chinese Academy of Sciences, Beijing, 100012, China}

\author{Xuelei Chen}
\affiliation{National Astronomical Observatories, Chinese Academy of Sciences, Beijing, 100012, China}

\author{Hongshi Zong}
\affiliation{Department of Physics, Nanjing University, Nanjing, 210093, China}
\affiliation{Joint Center for Particle, Nuclear Physics and Cosmology, Nanjing, 210093, China}

\begin{abstract}
It was proposed that 
the massive compact halo objects (MCHOs) 
would be produced during 
an earlier epoch of cosmology if 
the density perturbations 
are between $3 \times 10^{-4}$ and 0.3.
Then these objects can accrete 
dark matter particles onto them 
due to their high density. 
If the dark matter is in the form of 
the weakly interacting massive particles, 
the MCHOs can have a significant effect on 
the evolution of cosmology due to 
the dark matter annihilation within
them. Using the WMAP-7 years data, we 
investigated the constraints on the current abundance 
of MCHOs ($f_{MCHOs}$) formed during the $e^{+} e^{-}$ 
annihilation phase transition. We have found that the $2\sigma$ constraint 
is $f_{MCHOs} \lesssim 10^{-4}$ for dark matter masses
in the range between 1 GeV and 1 TeV.
\end{abstract}


\maketitle
\section {Introduction}
In order to form the present structure of cosmology, 
the initial density perturbations must satisfy the condition:
$\delta \sim 10^{-5}$. On the other hand, according to 
theory \cite{9704251}, if the density perturbations are larger 
than 0.3, primordial black holes (PBHs)  
can be formed during the radiation 
dominated era and existed if the mass is large enough \cite{0805.1531}.  
However, what about the result when it lies between 
these two cases?
In paper \cite{0908.0735}, the authors argued that 
if the density perturbations are between 
$3 \times 10^{-4}$ and 0.3, the nonbaryonic ultracompact minihalos
 named massive compact halo objects (MCHOs) can be formed 
during earlier times. 
Although within the conventional cases
the density perturbations are not 
larger enough to form these objects, 
they could be enhanced through the inflation potential 
or during the phase transitions in the early Universe \cite{9606125}. 
Because of their high density, these objects can accrete 
dark matter particles onto them.

Although the presence of dark matter in the Universe 
has been shown by many astrophysical observations, 
its nature still remains unknown. 
Among many models, 
the weakly interacting massive particles (WIMPs) model 
has been researched frequently \cite{Kamionkowski95,Bertone,Bergstrom09}.
According to the theory, WIMPs can annihilate into 
the standard particles such as electrons, positrons, 
protons, antiprotons, or photons. 
The authors of \cite{0908.4082} calculated 
the gamma-rays from these MCHOs which have been formed 
during three phase transitions: electroweak symmetry breaking,
QCD confinement and $e^{+}e^{-}$ annihilation. 
And in the last case, they found
that the integrated intensity of 
gamma-ray flux within the 100pc has exceeded 
the threshold of EGRET and Fermi-LAT. 
In \cite{1006.4970},using the Fermi gamma-ray 
observations, the authors investigated 
the constraints on the current abundance 
of MCHOs. 
They found that for the mass of MCHOs $\sim 10^{3} M_{\odot}$, 
the fraction is $\sim 10^{-7}$.

As discussed in papers \cite{zl}, 
the dark matter annihilation has a significant 
effect on the evolution of cosmology especially 
when the structure formation process is included. 
The structure formation starts
in the lower redshift $\sim 100$ while the MCHOs 
can be formed and accrete the dark matter particles
in the higher redshift $\sim 3000$. 
So the dark matter annihilation within them 
has an effect on the ionization and recombination 
before the structure formation begins. 
In this paper, we consider this
effect and try to give the constraints on the current abundance 
of MCHOs using the CMB observations.

One point to be noticed is that 
the PBHs can also accrete  
the dark matter particles to form similar objects
and moreover the PBHs themselves 
can emit photons, electrons and so on. 
So all of these objects can 
have effects on the evolution of cosmology
 \cite{Ricotti}. 
However, in this paper, these effects will not be considered.

This paper is organized as follows: 
We give equations and discuss 
how MCHOs affect the cosmological evolution 
in Sec. II. In Sec. III, 
we give our results of constraints on the current 
abundance of MCHOs using the WMAP data. 
We conclude in Sec. IV.

\section {the effects of MCHOs on cosmological evolution} 

Massive compact halos would be produced 
during the radiation domination epoch 
if the $\delta$ satisfies the condition: 
$3 \times 10^{-4} < \delta < 0.3$.
These objects can accrete 
dark matter particles by radial infall and then 
the mass of MCHOs evolves as \cite{0908.4082}:

\begin{equation}
\label{Mh}
M_\mathrm{MCHOs}(z) = \delta m \left(\frac{1 + z_\mathrm{eq}}{1+z}\right),
\end{equation}

where the $\delta m$ is the mass contained within a perturbation 
at the redshift of matter-radiation equality $z_\mathrm{eq}$. 
Following \cite{0908.4082}, we adopt 
$\delta m = 5.6 \times 10^{-19} M_\odot, 
1.1 \times 10^{-9} M_\odot$, and $0.33 M_\odot$ 
for three phase transitions: 
electroweak symmetry breaking (EW),
QCD confinement, and $e^{+}e^{-}$ annihilation.
 
The density profile of MCHOs is \cite{0908.4082}
\begin{equation}
\label{density}
\rho_{MCHOs}(r,z) = \frac{3f_\chi M_\mathrm{MCHOs}(z)}{16\pi R_\mathrm{MCHOs}(z)^\frac{3}{4}r^\frac{9}{4}},
\end{equation}

here ${R_\mathrm{NACHOs}(z)} = 
0.019\left(\frac{1000}{z+1}\right)\left(\frac{M_\mathrm{MCHOs}(z)}
{\mathrm{M}_\odot}\right)^\frac{1}{3} \mathrm{pc}$ 
and $f_{\chi}$ is the dark matter fraction.
We also accept the assumption that MCHOs 
stop growing at $z \approx 10$ because 
the structure formation process prevents further accretion 
after the redshift. 

In this paper, we assumed that 
the MCHOs have a monochromatic mass function, 
which means all of the MCHOs have the same mass, 
similar to the PBHs case \cite{0912.5297}.
We suppose that the abundance of MCHOs is the same 
everywhere and they do not merger with others. 
We neglect 
the energy loss of the dark matter 
annihilation production within MCHOs \cite{1003.3466}. 
Based on these assumptions, we can get the annihilation rate of MCHOs:

\begin{eqnarray}
\label{eq:fai}
\Gamma = N_{MCHOs} \Gamma' = 
N_{MCHOs} \frac{\langle\sigma v\rangle}{m_\chi^2} \int 4\pi r^2\rho^2(r,z)\mathrm{d}r \nonumber\\
= \frac{\rho_{0,MCHOs}}{M_{MCHOs}(z=0)}(1+z)^3 
\frac{\langle\sigma v\rangle}{m_\chi^2} \int 4\pi r^2\rho^2(r,z)\mathrm{d}r \nonumber\\
= \frac{f_{MCHOs} \rho_{0,critical}}{M_{MCHOs}(z=0)}(1+z)^3 
\frac{\langle\sigma v\rangle}{m_\chi^2} \int 4\pi r^2\rho^2(r,z)\mathrm{d}r.
\end{eqnarray}

where $\Gamma'$ is the annihilation rate within one of the MCHOs 
and $\Gamma$ is the annihilation rate per unit volume of the MCHOs. 
$N_{MCHOs}$ is the number density of MCHOs, 
$f_{MCHOs}$ = $\rho_{MCHOs}/\rho_{critical}$ and 
this definition is different from \cite{1006.4970}.  
The limitation of the integral is from $r_{cut}$ to $R_{MCHOs}$. 
The $r_{cut}$ is \cite{0908.4082,1006.4970,0207125}
 
\begin{eqnarray}
\label{eq:r_cut}
\rho(r_{cut}) = \frac{m_{\chi}}{\langle \sigma v \rangle(t_{0}-t_{i})}
\end{eqnarray}

where $t_{0} \approx 13.7 Gyr$ \cite{0908.4082,1006.4970} 
is the age of the Universe, 
$t_{i}$ is the time of the MCHOs formation and we choose 
$t_{i}(z_{eq}) \approx 77kyr$ also used by \cite{1006.4970}.

Besides the MCHOs, we consider 
the dark matter halos in this paper. 
Their contribution can be treated 
as the 'clumping factor' $C(z)$ relative 
to the smooth case \cite{cumberbatch}:

\begin{eqnarray}
\label{eq:cz}
C(z)&=&1 + {\Gamma_{halo}(z) \over \Gamma_{smooth}(z)}\nonumber\\
&&=1 + {(1 + z)^3 \over \bar \rho_{\rm DM}^2(z)}\int 
{\rm d}M\frac{{\rm d}n}{{\rm d}M}(M,z) \\
&&\times\int\rho^2(r)4\pi r^2{\rm d}r\nonumber 
\end{eqnarray}

where $\Gamma$ stands for the dark matter annihilation rate. 
$dn/dM$ is the halos mass function 
and we use the Press-Schechters formalism \cite{press} 
to do our calculations.  
On the other hand, through the simulation, 
it was found that  
there are many substructures in dark matter halos \cite{diemand}. 
These subhalos can also enhance the dark matter 
annihilation rate. In our paper, we  
include these subhalos, 
neglect the contributions 
from the sub-sub-, and 
use the smallest mass of them 
$\sim 10^{-6}M_\odot$ \cite{green,diemand}.  
We consider $\sim 10\%$ halos 
mass within the subhalos, 
use the power law 
form of mass function $\sim M^{-\beta}$ and adopt 
$\beta = 1.95$ \cite{diemand}. 
So the total clumping factor of dark matter 
halos and subhalos can be written as \cite{cumberbatch}: 

\begin{equation}
C_{taotal} = 1 + (C_{halos} - 1) + (C_{sunhalos} - 1)
\end{equation}

Considering the dark matter annihilation, 
the evolution of ionization fraction $x_{e}$ can be written as \cite{cxl,zl}:
\begin{equation}
(1+z){dx_e\over dz} = {1\over H(z)}[R_s(z)-I_s(z)-I_\chi(z)]
\end{equation}
where $R_s$ is the standard recombination rate, 
and $I_s$ is the ionization rate by standard sources, 
$I_\chi$ is the ionization rate sourced by dark matter 
which is given as \cite{zl}
\begin{equation}
{I_\chi} = {\chi_i f{2 m_\chi c^{2}\over n_b E_b}}{\Gamma_{total}} 
\end{equation}
where $n_b$ is the baryon number density and the $E_b = 13.6 eV$ 
is the ionization energy.
$\Gamma_{total}$ is the total dark matter
annihilation rate including the MCHOs and halos. 
$f$ that depends on the redshift and  
the production of dark matter annihilation \cite{slatyer}
is the released energy fraction 
depositing in the baryonic gas
during the annihilation. 
In this paper, we assume that the 
total energy released by the 
annihilation is deposited, which means $f = 1$. 
$\chi_i$ is the energy fraction which 
ionizes the baryonic gas and we  
accept the form given by \cite{cxl} 
\begin{equation}
{\chi_i} = {\left(1 - {x_e}\right) / 3}
\end{equation}  
where $x_e$ is the fraction of free electrons. 

Following the method in papers \cite{cxl} \cite{zl}, 
we modified the public code CAMB \cite{camb} 
in order to include the contributions from 
MCHOs and dark matter halos.

\begin{figure}
\epsfig{file=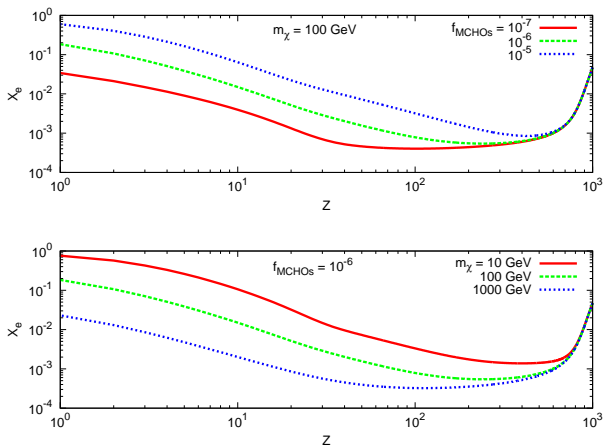,width=0.5\textwidth}
\caption{The change of the ionization fraction 
$x_{e}$ as a function of the redshift $z$. 
Upper panel: We fixed the mass of dark matter 
$m_{\chi} = 100 GeV$, and changed the current fraction 
of MCHOs $f_{MCHOs} = 10^{-7},10^{-6},and 10^{-5}$ from 
bottom to up. 
Lower panel: We have fixed the current fraction 
of MCHOs $f_{MCHOS} = 10^{-6}$ and 
changed the dark matter mass $m_{\chi} = 10,100,and 1000$ 
from top to bottom. Here we fixed the value
$\langle \sigma v \rangle = 3 \times 10^{-26} cm^{3} s^{-1}$.}
\label{fig:xe_z}
\end{figure}

\begin{figure}
\epsfig{file=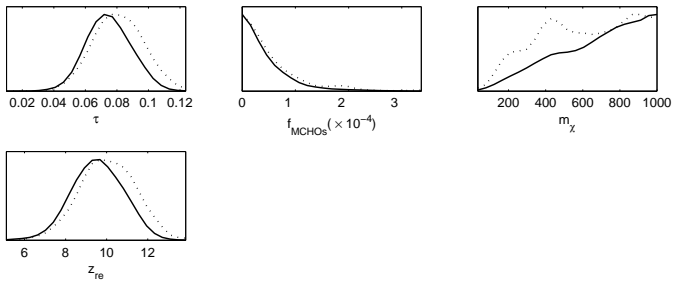,width=0.5\textwidth}
\caption{The marginalized probability distribution 
function of parameters $\tau, f_{MCHOs}, m_{\chi}, z_{re}$ (solid curve) 
and the relative mean likelihood (dotted curve). }
\label{fig:test_1D}
\end{figure}

\begin{figure}
\epsfig{file=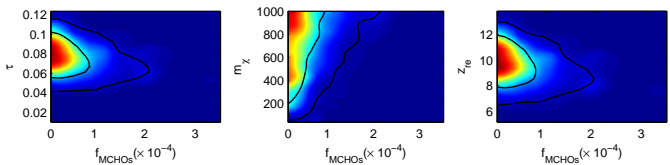,width=0.5\textwidth}
\caption{The 2D contours of the distribution of 
$f_{MCHOs}$ and the parameters $\tau, m_{\chi}$, and $z_{re}$. }
\label{fig:test_2D}
\end{figure}

In Fig. ~\ref{fig:xe_z}, 
we show the evolution of $x_{e}$ as a function of redshift z 
for the MCHOs 
formed during the $e^{+}e^{-}$ annihilation phase transition. 
For the cosmological parameters, 
we use the WMAP7 results \cite{wmap7}. 
We can see that the larger the fraction 
of MCHOs, the more obvious change of the ionization fraction 
can be made.
We also notice that the influence of dark matter annihilation within 
MCHOs on the cosmological 
ionization fraction is similar to the cases of dark matter decay \cite{cxl} 
or the PBHs \cite{0805.1531} for $M_{PHB} > 10^{14} kg$. 

For the other MCHOs formed 
during the EW and 
QCD phase transitions, 
the contribution 
on the cosmological evolution is much weaker 
due to their much smaller mass. 
For this property, it can be 
seen from the paper \cite{0908.4082} where 
the integrated gamma-ray flux above 100MeV for EW and QCD cases 
are $\sim$ 12 and 6 orders lower than the $e^{+}e^{-}$ case respectively. 
So in this paper, we do not consider 
these two cases. 

\section{Constraints from WMAP data} 

In this paper,
we have modified the public 
COSMOMC code \cite{cosmomc} and used the 7-year WMAP 
results, both the temperature and polarization data, 
to get the constraints of related parameters. 
We consider 6 cosmological parameters: 
$\Omega_b h^2,\Omega_d h^2,\theta,\tau,n_s$,and $A_s$, 
where $\Omega_b h^2$ and $\Omega_d h^2$  are the density of 
baryon and dark matter, $\theta$ is the ratio of
the sound horizon at recombination to its angular diameter distance 
multiplied by 100, $\tau$ is the optical depth, 
and $n_s$ and $A_s$ are the the spectral index and amplitude of
the primordial density perturbation power spectrum. 
We fix the value $\langle \sigma v \rangle$ = 
$3 \times 10^{-26} \rm cm^{3} \rm s^{-1}$ 
and treat $m_{\chi}$ and $f_{MCHOs}$ as free parameters.
For recent observations \cite{bernabei,adriani,chang}, 
the preferred dark matter mass spreads from several 
GeV to TeV. So in our paper, we 
set the change range of dark matter mass from 1GeV to 1TeV 
and the prior of the $f_{MCHOS}:[0,10^{-2}]$. 
For the convergence diagnostic of MCMC, we use 
the default value of the Gelman and Rubin statistics in COSMOMC
([variance of chain means]/[mean of chain variances]): R-1 = 0.03. 
The results are shown in Table ~\ref{tab:machos} 
where some important parameters
such as the optical depth ($\tau$), 
the redshift of reionization ($z_{re}$), 
the current abundance of MCHOs ($f_{MCHOs}$) 
and the dark matter mass ($m_{\chi}$) are given. 

\begin{table*}
\caption{\label{tab:machos}
 Posterior constraints on the fractions of MCHOs, 
the mass of dark matter and the related cosmological parameters. }.\\
\begin{center}
\begin{tabular}{|c|c|c|c|c|}
\hline
 Parameter &$\tau$&$z_{re}$&$f_{MCHOs} (10^{-4})$&$m_{\chi} (GeV)$ \\
\hline
  Mean  &$0.0753$&$9.59$&$0.48$&$682.0$  \\
\hline
  $2\sigma$ lower  &$0.0538$&$7.09$&$0.0$&$188.4$   \\ 
\hline
  $2\sigma$ upper &$0.0983$&$11.95$&$1.52$&$999.3$  \\ 
\hline

\end{tabular}
\end{center}
\end{table*}

It can be seen that the $2\sigma$ limitation of the MCHOs fraction 
is $f_{MCHOS} \sim 10^{-4}$. 
As shown above, the larger the mass of dark matter, 
the smaller the contributions are. So the $2\sigma$ constraints 
on the dark matter mass approach our prior limitation. 
We also plot the 1D and 2D probability distribution 
in Fig. ~\ref{fig:test_1D} and ~\ref{fig:test_2D}. 

From Fig. ~\ref{fig:test_1D}, we see 
that $f_{MCHOs}$ has the biggest probability 
at zero and decreases approximately to zero 
at $\sim 3 \times 10^{-4}$. 
For the dark matter mass, the constraints are 
weaker, which is much more obviously in 
the 2D probability 
distribution of parameters $f_{MCHOs}$ 
and $m_{\chi}$ where the dark matter mass 
spreads a larger range.

\section {Conclusion} 

We have investigated the current 
abundance of MCHOs formed 
in an earlier epoch due to 
the large density perturbations ($3 \times 10^{-4} <\delta< 0.3$)
using the WMAP-7 years data.
We found that for the mass range 1GeV $\sim$ 1TeV, 
the current abundance of MCHOs which are produced during 
the $e^{+}e^{-}$ annihilation phase transition is
 $f_{MCHOs} \sim 10^{-4}$ 
corresponding to the $2\sigma$ limitation. This is the first constraint 
using the WMAP data and it is comparable with 
the results in paper \cite{1006.4970}, 
where they used the Fermi gamma-ray observations. 

The influences of MCHOs 
formed during the electroweak breaking 
and QCD confinement phase transitions 
on the cosmological evolution 
are much weaker than the $e^{+}e^{-}$ annihilation  
case because of their much smaller mass, 
so the WMAP cannot give 
the significative constraints 
on their current abundance. 
 
We also find that the effect on the cosmological ionization fraction of 
dark matter annihilation within 
MCHOs is similar to the cases of dark matter decay \cite{cxl} or  
the PBHs for $M_{PHB} >10^{14}kg$ \cite{0805.1531}.

\section{Notes}

\par

THE FIRST NOTE : We also have used the $\gamma$-ray data obtained 
by the Fermi observations to get the constrains on them \cite{gamma}.

\par

THE SECOND NOTE: Following the suggestions from Pat Scott, 
we found an error for the choice of values of $\delta m$ in Sec. II. 
So we corrected the values according to \cite{error} 
and recalculated. We found that 
the results are not changed for our model. 
Because the integration of Eq.(3) 
is decreased by a factor of $\sim 10^{-4}$ (see also \cite{error}). 
But the $M_{MCHO(z=0)}$ is also changed 
by a same factor (0.33/390 $\sim 10^{-4}$). 
\par

We notice that the similar work was also 
done by D. Zhang \cite{zhang}, and they disscussed 
the much more general case. They also investigated 
the accretion of gas by MCHOs and other things.
We are glad to suggest that some ones who are interested in 
MCHOs can refer to the Zhang's paper for 
the detailed discussion.

\section{Acknowledgments}

1: We thank Lei Feng for helpful discussions. 
Yupeng yang thanks Yan Qian for 
improving the manuscript. 
Our MCMC chains computation was performed on 
the Shenteng 7000 system of the Supercomputing 
Center of the Chinese Academy of Sciences. 
\par
2: We thank Pat Scott very much for comment 
and suggestion.

\newcommand\PR[3]{~Phys. Rept.{\bf ~#1}, #2~(#3)}
\newcommand\NJP[3]{~New J.Phys.{\bf ~#1}, #2~(#3)}
\newcommand\PRD[3]{~Phys. Rev. D{\bf ~#1}, #2~(#3)}
\newcommand\APJ[3]{~Astrophys. J.{\bf ~#1}, #2~ (#3)}
\newcommand\PRL[3]{~Phys. Rev. Lett.{\bf ~#1}, #2~(#3)}
\newcommand\APJS[3]{~Astron. J. Suppl.{\bf ~#1}, #2~(#3)}
\newcommand\EPJC[3]{~Eur. Phy. J. C{\bf ~#1}, #2~(#3)}
\newcommand\Nature[3]{~Nature{\bf ~#1}, #2~(#3)}
\newcommand\JCAP[3]{~JCAP{\bf ~#1}, #2~(#3)}
\newcommand\APJL[3]{~Astrophys. J. Lett.{\bf ~#1}, #2~ (#3)}

\end{document}